\documentclass[prb,superscriptaddress,showpacs,floatfix,tightenlines,10pt,twocolumn]{revtex4}
\usepackage{amssymb}
\usepackage{amsmath}
\usepackage{graphicx}

\begin{document}

\title{Ising quantum Hall ferromagnetism in Landau levels $\left\vert
N\right\vert \geq 1$ of bilayer graphene}
\author{Wenchen Luo}
\affiliation{D\'{e}partement de physique, Universit\'{e} de Sherbrooke, Sherbrooke, Qu%
\'{e}bec, J1K 2R1, Canada}
\author{R. C\^{o}t\'{e}}
\affiliation{D\'{e}partement de physique, Universit\'{e} de Sherbrooke, Sherbrooke, Qu%
\'{e}bec, J1K 2R1, Canada}
\author{Alexandre B\'{e}dard-Vall\'{e}e}
\affiliation{D\'{e}partement de physique, Universit\'{e} de Sherbrooke, Sherbrooke, Qu%
\'{e}bec, J1K 2R1, Canada}

\begin{abstract}
A magnetic field applied perpendicularly to the chiral two-dimensional
electron gas (C2DEG)\ in a Bernal-stacked bilayer graphene quantizes the
kinetic energy into a discrete set of Landau levels $N=0,\pm 1,\pm 2,...$%
While Landau level $N=0$ is eighfold degenerate, higher Landau levels ($%
\left\vert N\right\vert \geq 1$) are fourfold degenerate when counting spin
and valley degrees of freedom. In this work, the Hartree-Fock approximation
is used to study the phase diagram of the C2DEG at integer fillings $%
\widetilde{\nu }=1,2,3$ of these higher Landau levels. At these filling
factors, the C2DEG is a valley or spin Ising quantum Hall ferromagnet. At
odd fillings, the C2DEG is spin polarized and has all its electrons in one
valley or the other. There is no intervalley coherence in contrast with most
of the the ground states in Landau level $N=0.$ At even filling, $\widetilde{%
\nu }=2,$ the C2DEG is either fully spin polarized with electrons occupying
both valleys or spin unpolarized with electron occupying one of the two
valleys. A finite electric field (or bias) applied perpendicularly to the
plane of the C2DEG induces a series of first-order phase transitions between
these different ground states. The transport gap or its slope is
discontinuous at the bias where a transition occurs. Such discontinuity may
result in a change in the transport properties of the C2DEG at that bias.
\end{abstract}

\date{\today }
\maketitle

\section{INTRODUCTION}

In a strong perpendicular magnetic field $\mathbf{B}$, the kinetic energy of
a two-dimensional electron gas (2DEG) is quantized into a series of discrete
Landau levels with a macroscopic degeneracy $N_{\varphi }=SB/\Phi _{0},$
where $S$ is the sample area and $\Phi _{0}$ is the magnetic flux quantum.
This quantization leads to the integer quantum Hall effect which has been
the subject of intense study over the past $35$ years\cite{Review1,Review2}.
The ground state of the 2DEG at odd values of the filling factors $\nu
=N_{e}/N_{\varphi }$ (where $N_{e}$ is the number of electrons added to the
2DEG at the neutrality point) is completely spin polarized even when the
Zeeman coupling is set to zero because a parallel orientation of the spins
minimizes the Coulomb repulsion. This ground state is referred to as a spin
quantum Hall ferromagnet (spin-QHF). In the absence of Zeeman coupling, this
state has a SU(2) symmetry in spin space.

In a double quantum well system (DQWS), each electron has an extra layer
degree of freedom which can be associated with a layer pseudospin $P_{z}=\pm
1/2$. At filling factor $\nu =1$ and when the 2DEG is spin polarized, the
minimization of the capacitive energy of the 2DEG forces all electrons to be
in a symmetric combination of the right and left layer in the ground state.
The layer pseudospins are thus all aligned in the $xy-$plane in pseudospin
space and the 2DEG is this time referred to as a layer-QHF\cite{Kmoon}. In
the absence of tunneling between the two layers, the layer-QHF has a U(1)
symmetry which is associated with the invariance of the ground state energy
with respect to the orientation of the layer-pseudospin in the $xy-$plane.
The physics of the layer-QHF is reviewed in Ref. \onlinecite{Ezawa}.

Quantum Hall ferromagnetism can be associated with many other type of
degrees of freedom. In graphene, for example, the ground state of the chiral
2DEG (C2DEG) at $1/4$ and $3/4$ fillings of the Landau levels $\left\vert
N\right\vert \geq 1$ is a valley-QHF where the two pseudospin states are now
associated with the non-equivalent valleys $K_{+}$ and $K_{-}.$ To a good
approximation, the Coulomb interaction is independent of the valley index%
\cite{Goerbig} and since there is no symmetry-breaking term associated with
the valley degree of freedom, the Hamiltonian has a SU(2) symmetry in
valley-pseudospin space. Such is also the case in Landau level $N=0$ of a
Bernal-stacked bilayer graphene (BLG), a system which has been extensively
studied both theoretically and experimentally over the past few years\cite%
{BarlasRevue2012}. Because of its extra \textit{orbital} degeneracy, the $%
N=0 $ Landau level in BLG has an eightfold degeneracy in the absence of
Zeeman coupling in the minimal tight-binding model where only the in-plane, $%
\gamma _{0},$ and inter-plane, $\gamma _{1},$ hopping terms are considered.
In contrast with graphene, an electric field (or \textit{bias})\ applied
perpendicularly to the plane of the layers in BLG breaks the valley
degeneracy and also the layer degeneracy since valley and layer degrees of
freedom are equivalent in $N=0$. Moreover, when Coulomb interaction is taken
into account, a rich set of quantum\ Hall ferromagnets emerges at integer
filling factors $\nu \in \left[ -4,4\right] $ as the bias is varied\cite%
{Lambert}.

In this paper, the Hartree-Fock approximation (HFA) is used to study the
quantum Hall ferromagnetic ground states of the C2DEG in a Bernal-stacked
graphene bilayer in the higher Landau levels $\left\vert N\right\vert \geq
1. $ By contrast with $N=0,$ the higher Landau levels are fourfold
degenerate in the absence of Zeeman coupling because there is no orbital
degree of freedom in higher Landau levels. The QHF\ states are studied at
integer fillings $\widetilde{\nu }=1,2,3$ of the Landau level $N\ $and as a
function of the magnetic field strength $B$ and the electrical bias $\Delta
_{B}$. The behavior of the C2DEG in $\left\vert N\right\vert \geq 1$ is
found to be very different than that in level $N=0.$ In the latter case, the
ground states at zero bias, with the exception of $\nu =0,$ are valley-QHFs
with a U(1) symmetry in the $xy-$plane. In the former case, the quantum Hall
ferromagnetism is of the Ising type with two degenerate ground states at $%
\Delta _{B}=0$. No intervalley coherence is possible. At finite Zeeman
coupling and in the absence of bias, the ground states at $\widetilde{\nu }%
=1,3$ are Ising valley-QHFs with a $Z_{2}$ symmetry in valley-pseudospin
space. A finite bias $\Delta _{B}$ can induce a first order phase transition
between the two pseudospin states. At filling $\widetilde{\nu }=2$ of level $%
N$, the C2DEG is a spin-QHF below a certain critical bias $\Delta
_{B}^{\left( c\right) }$ that depends on the magnetic field and on the
dielectric constant of the substrate. Above $\Delta _{B}^{\left( c\right) }$%
, the system is a valley-QHF with valley pseudospin $P_{z}=\pm 1.$ The phase
diagram is found to depend sensitively on the Landau level index $N$ and on
the value of the dielectric constant $\kappa .$ The transition between two
QHF\ phases is accompanied by a discontinuity in the Hartree-Fock
electron-hole gap (the transport gap) or its slope. Depending on the Landau
level broadening due to disorder, this discontinuity may lead to a
disappearance of the quantum Hall effect and to an increase in the
longitudinal resistivity at the transition. Such an effect has been seen in
a recent experiment\cite{Tutuc} on double bilayer graphene in Landau levels $%
N\leq 0$. We find that our numerical results for the behavior of $\Delta
_{B}^{\left( c\right) }$ with magnetic field are in qualitative agreement
with these experimental results .

There has been up to now very few studies of the phase diagram of the C2DEG
in higher Landau levels of BLG. Wigner and Skyrme crystal states have been
studied near integer filling factors\cite{Sakurai}, but the Ising behavior
reported in the present article has not been discussed before in this
system. It has, however, been studied previously in many other systems.
Usually, the Ising behavior occurs when any two different Landau levels
simultaneously approach the chemical potential. At the crossing point, the
nature of the ground state is sensitive to the microscopic character of the
crossing Landau levels and different types of QHF can occur. At even integer
filling factor, the crossing often gives rise to a first-order paramagnetic
to ferromagnetic transition\cite{Quinn,Daneshvar}. A classification scheme
that applies to single layer and bilayer semiconductor 2DEGs is presented in
Ref. \onlinecite{Classification}. In the present work, the Ising behavior
can be related to the crossing of two sub-levels in a Landau level $N$ or be
exchange-energy driven and not related to any Landau level crossings.

This paper is organized in the following way. Sections II, III and IV
present the tight-binding model of BLG, the Hartree-Fock approximation for
the Coulomb interaction and the Green's function method used to calculate
the order parameters of the different phases. Section V presents the
pseudospin language used to describe the various phases. The phase diagrams
for different filling factors are presented in Secs. VI,VII,VIII for $%
\widetilde{\nu }=1,3,2$ respectively. Section IX contains a discussion of
our results and a comparison with the available experimental data.

\section{NON-INTERACTING\ HAMILTONIAN IN LANDAU LEVELS $N\geq 1$}

The system considered in this paper is a Bernal-stacked graphene bilayer in
a transverse magnetic $\mathbf{B=}B\widehat{\mathbf{z}}$ and electric field $%
\mathbf{E.}$ The electric field induces a potential difference (hereafter
called the \textit{bias}) $\Delta _{B}=Ed$ between the two layers, where $%
d=3.337$ \AA\ is the interlayer separation. The crystal structure of each
graphene layer is a honeycomb lattice that can be described as a triangular
Bravais lattice with a basis of two carbon atoms $A_{n}$ and $B_{n},$ where $%
n=1,2$ is the layer index. The triangular lattice constant $%
a_{0}=2.\,\allowbreak 46$ \AA\ $=\sqrt{3}c,$\ where $c=1.42$ \AA\ is the
distance between two adjacent carbon atoms. The Brillouin zone of the
reciprocal lattice is hexagonal and has two nonequivalent valley points $%
K_{\xi }=\left( \frac{2\pi }{a_{0}}\right) \left( \xi \frac{2}{3},0\right) ,$
where $\xi =\pm $ $1$ is the valley index\cite{ReviewGraphene}.

In the absence of magnetic field and bias, the electronic band structure
consists of four bands. The two middle bands meet at the six valley points
while the two high-energy bands are separated by a gap $\gamma _{1}$ from
the two middle, low-energy bands (see, for example, Fig. 1 of Ref. %
\onlinecite{Manuel}). In a finite magnetic field, each band is split into a
set of Landau levels. Below, we use the index $j=1,2,3,4$ to refer to the
set of Landau levels that originate from each band. The bands are indexed in
order of increasing energy.

In the continuum approximation, the tight-binding Hamiltonian for $\mathbf{B}%
=0$ is expanded to linear order in the wave vector $\mathbf{p}=\mathbf{k}-%
\mathbf{K}_{\pm }$ in each valley. The effect of a magnetic field is taken
into account by making the Peierls substitution $\mathbf{p}\rightarrow 
\mathbf{P}=\mathbf{p}+e\mathbf{A}/\hslash c$ (with $e>0$ for an electron)
for the wave vector where $\mathbf{A}=\left( 0,Bx,0\right) $ is the vector
potential in the Landau gauge. The tight-binding Hamiltonian in the basis $%
\left\{ A_{1},B_{1},A_{2},B_{2}\right\} $ for valley $K_{-}$ and $\left\{
B_{2},A_{2},B_{1},A_{1}\right\} $ for valley $K_{+}$ is given by 
\begin{equation}
H_{\xi }^{\left( 0\right) }=\left( 
\begin{array}{cccc}
\delta _{0}-\xi \frac{\Delta _{B}}{2} & \xi \alpha _{0}a & \xi \alpha
_{4}a^{\dag } & -\gamma _{1} \\ 
\xi \alpha _{0}a^{\dag } & -\xi \frac{\Delta _{B}}{2} & \xi \alpha _{3}a & 
\xi \alpha _{4}a^{\dag } \\ 
\xi \alpha _{4}a & \xi \alpha _{3}a^{\dag } & \xi \frac{\Delta _{B}}{2} & 
\xi \alpha _{0}a \\ 
-\gamma _{1} & \xi \alpha _{4}a & \xi \alpha _{0}a^{\dag } & \delta _{0}+\xi 
\frac{\Delta _{B}}{2}%
\end{array}%
\right) ,  \label{t3}
\end{equation}%
where the parameter%
\begin{equation}
\alpha _{i}=\sqrt{\frac{3}{2}}\frac{a_{0}}{\ell }\gamma _{i},  \label{t1}
\end{equation}%
with $\ell =\sqrt{\hslash c/eB}$ the magnetic length. The hopping parameters
in $H_{\xi }^{\left( 0\right) }$are: $\gamma _{0}$ the in-plane
nearest-neighbor (NN) hopping; $\gamma _{1}$ the interlayer hopping between
carbon atoms that are immediately above one another (i.e. $A_{1}-B_{2}$); $%
\gamma _{3}$ the interlayer NN hopping term between carbon atoms of
different sublattices (i.e. $A_{2}-B_{1}$) and $\gamma _{4}$ the interlayer
next NN hopping term between carbons atoms in the same sublattice (i.e. $%
A_{1}-A_{2}$ and $B_{1}-B_{2}$). The parameter $\delta _{0}$ represents the
difference in the crystal field between sites $A_{1},B_{2}$ and $A_{2},B_{1}$
and $\Delta _{B}$ is the potential difference between the two layers. For
all calculations done in this paper, we use for the value of each parameter%
\cite{Values}: $\gamma _{0}=3.12$ eV, $\gamma _{1}=0.39$ eV, $\gamma
_{3}=0.29,$ $\gamma _{4}=0.12$ eV and $\delta _{0}=0.0156$ eV. The ladder
operators $a,a^{\dag }$ are defined such that $a\varphi _{n}\left( x\right)
=-i\sqrt{n}\varphi _{n-1}\left( x\right) $ and $a^{\dag }\varphi _{n}\left(
x\right) =i\sqrt{n+1}\varphi _{n+1}\left( x\right) ,$ where $\varphi
_{n}\left( x\right) $ with $n=0,1,2,...$ are the eigenfunctions of the
one-dimensional harmonic oscillator that enter in the definition of the
Landau-gauge wave functions%
\begin{equation}
h_{n,X}\left( \mathbf{r}\right) =\frac{1}{\sqrt{L_{y}}}e^{-iXy/\ell
^{2}}\varphi _{n}\left( x-X\right) ,
\end{equation}%
where $X$ is the guiding-center quantum number.

\begin{figure}[tbph]
\includegraphics[scale=0.8]{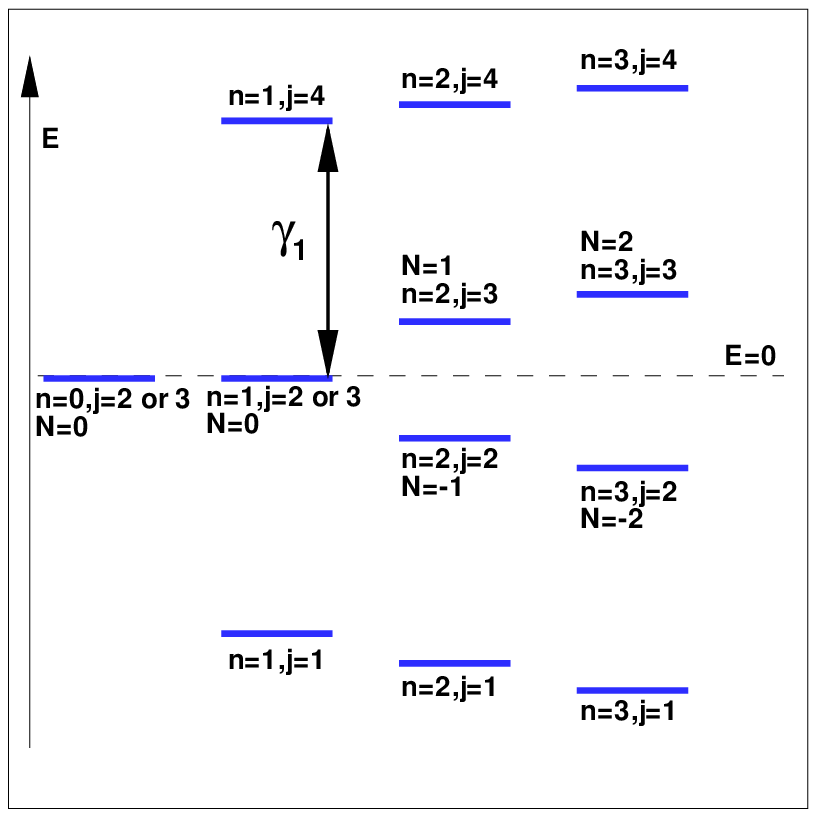}
\caption{(Color online) Labeling of the energy levels for one valley and one
spin component at zero bias and for $\protect\gamma _{4}=\protect\delta %
_{0}=0$. The vertical position of each line does not reflect the actual
energy of each level.}
\label{figure1}
\end{figure}

When the warping term $\gamma _{3}=0,$ the eigenspinors of $H_{\xi }^{\left(
0\right) }$ in the basis $\left\{ A_{1},B_{1},A_{2},B_{2}\right\} $ for both 
$K_{+}$ and $K_{-}$ have the form%
\begin{equation}
\psi _{\xi =-1,n,j,X}^{\left( 0\right) }\left( \mathbf{u}\right) =\left( 
\begin{array}{c}
c_{-,n,j,1}h_{n-1,X}\left( \mathbf{r}\right) \chi _{1}\left( z\right) \\ 
c_{-,n,j,2}h_{n,X}\left( \mathbf{r}\right) \chi _{1}\left( z\right) \\ 
c_{-,n,j,3}h_{n-2,X}\left( \mathbf{r}\right) \chi _{2}\left( z\right) \\ 
c_{-,n,j,4}h_{n-1,X}\left( \mathbf{r}\right) \chi _{2}\left( z\right)%
\end{array}%
\right) ,  \label{psi1}
\end{equation}%
and%
\begin{equation}
\psi _{\xi =+1,n,j,X}^{\left( 0\right) }\left( \mathbf{u}\right) =\left( 
\begin{array}{c}
c_{+,n,j,4}h_{n-1,X}\left( \mathbf{r}\right) \chi _{1}\left( z\right) \\ 
c_{+,n,j,3}h_{n-2,X}\left( \mathbf{r}\right) \chi _{1}\left( z\right) \\ 
c_{+,n,j,2}h_{n,X}\left( \mathbf{r}\right) \chi _{2}\left( z\right) \\ 
c_{+,n,j,1}h_{n-1,X}\left( \mathbf{r}\right) \chi _{2}\left( z\right)%
\end{array}%
\right) ,  \label{psi2}
\end{equation}%
where it is understood that $h_{n,X}\left( \mathbf{r}\right) =0$ if $n<0$
and the function $\left\vert \chi _{i}\left( z\right) \right\vert
^{2}=\delta \left( z-z_{i}\right) ,$ where $z_{i}$ with $i=1,2$ is the
position of layer $i$ along the $z$ axis and $\left\vert
z_{2}-z_{1}\right\vert =d$. The three-dimensional vector $\mathbf{u}=\left( 
\mathbf{r},z\right) ,$ where $\mathbf{r}$ is a two-dimensional vector in the
plane of the bilayer. The corresponding energy levels are written as $E_{\xi
,n,j}^{\left( 0\right) }.$ They are independent of $X$ and so they have
degeneracy $N_{\varphi }=S/2\pi \ell ^{2},$ where $S$ is the 2DEG area. As
Fig. \ref{figure1} shows, for $n=0$ there is one eigenspinor that belongs to
the Landau levels of band $j=2$ for $K_{+}$ or $j=3$ for $K_{-}$. For $n=1,$
there are three eigenspinors belonging to bands $j=1,2,4$ for $K_{+}$ and to 
$j=1,3,4$ for $K_{-}$. The solutions $n=0,j=2$ or $3$ and $n=1,j=2$ or $3$
are degenerate when $\Delta _{B},\gamma _{4},\delta _{0}=0$ and are
considered as belonging to Landau level $N=0$ which has thus an eightfold
degeneracy when counting spin and valley degrees of freedom. This degeneracy
is called the \textit{orbital} degeneracy. For $n\geq 2,$ there are four
energy eigenspinors, one for each energy band $j$. Hereafter, we only work
with bands $2$ and $3$ and so we label the levels of band $j=2$ by $%
N=-1,-2,-3,...$ and those of band $j=3$ by $N=1,2,3,...$ as indicated in
Fig. \ref{figure1}. All Landau levels $\left\vert N\right\vert >0$ are
fourfold degenerate when spin and valley degrees of freedom are considered.

The warping term $\gamma _{3}$ couples the eigenspinors together so that a
solution of the full non-interacting Hamiltonian can be written as the
linear combination 
\begin{equation}
\widetilde{\psi }_{\xi ,n,j,X}^{\left( 0\right) }\left( \mathbf{u}\right)
=\sum_{n^{\prime },j^{\prime }}b_{\xi ,n^{\prime },j^{\prime }}^{\left(
0\right) }\psi _{\xi ,n^{\prime },j^{\prime },X}\left( \mathbf{u}\right)
\label{psibilayer}
\end{equation}%
with the normalization condition $\sum_{n,j}\left\vert b_{\xi
,n,j}\right\vert ^{2}=1$. In this paper, we are interested in the phase
diagram of the 2DEG in Landau levels $\left\vert N\right\vert =1,2,3.$ In
order to estimate the importance of the hopping term on these levels, we
compare the energies $E_{\xi ,n,j}^{\left( 0\right) }$ and coefficients $%
b_{\xi ,n,j}$ computed with or without it. The results are shown in Table %
\ref{Table1} for the valley $\xi =+1$, band $j=3,$ and for different values
of the magnetic field at zero bias. Clearly, for these levels, neglecting
the warping term $\gamma _{3}$ is a good approximation. That is why,
hereafter, we set $\gamma _{3}=0$ and use the simpler eigenspinors given by
Eqs. (\ref{psi1}),(\ref{psi2}). We remark that we do not use the effective
two-component model\cite{McCann} in this paper since it is not good at
describing the higher Landau levels\cite{Manuel}.

\begin{center}
\begin{table}[tbp] \centering%
\begin{tabular}{|l|l|l|l|l|}
\hline
$N$ & $B$ (T) & $\gamma _{3}=0,E_{_{+,N+1,3}}^{\left( 0\right) }$ & $\gamma
_{3}\neq 0,E_{_{+,N+1,3}}^{\left( 0\right) }$ & $\left\vert
b_{+,N+1,3}\right\vert ^{2}$ \\ \hline
$1$ & $10$ & $0.04857$ & $0.04821$ & $0.988$ \\ 
$2$ & $10$ & $0.07847$ & $0.07845$ & $0.975$ \\ 
$3$ & $10$ & $0.105$ & $0.105$ & $0.970$ \\ \hline
$1$ & $40$ & $0.153$ & $0.153$ & $0.995$ \\ \hline
$2$ & $40$ & $0.232$ & $0.232$ & $0.989$ \\ \hline
$3$ & $40$ & $0.298$ & $0.297$ & $0.983$ \\ \hline
\end{tabular}%
\caption{Non-interacting energies and expansion coefficients calculated with and without the warping term for different values of the Landau level $N,j=3$ at magnetic fields
$B=10$ T and $40$ T in the valley $K_+$.}\label{Table1}%
\end{table}%
\end{center}

\begin{figure}[tbph]
\includegraphics[scale=0.8]{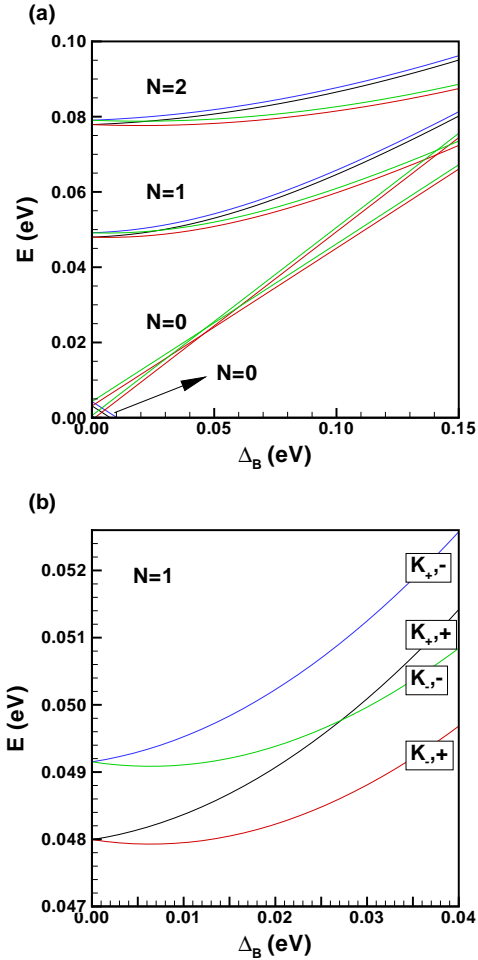}
\caption{(Color online) (a) Dispersion of the Landau levels $N=0,1,2$ with
bias $\Delta _{B}$ for $B=10$ T; (b) part of the same energy spectrum for
level $N=1$ showing the crossing between the sub-levels $(K_{+},+)$ and $%
(K_{-},-).$}
\label{figure2}
\end{figure}

Figure \ref{figure2} (a) shows the energy of the Landau levels $N=0,1,2$ as
a function of the bias $\Delta _{B}$ for $B=10$ T. The identification of the
levels is made in Fig. \ref{figure2} (b). For Landau level $N=0,$ the levels
shown are those corresponding to $n=0,1$ (of both spins) in valley $K_{-}.$
By contrast, the energy of levels $n=0,1$ in valley $K_{+}$ (not shown)\
decreases with bias. It is important to notice that, in Landau level $N=0,$
valley and layer indices are equivalent\ (in the two-component model) but
not this is not true in higher Landau levels as can easily be seen from the
eigenspinors in Eqs. (\ref{psi1}),(\ref{psi2}).

Crossings between different Landau levels occur for $\Delta _{B}\gtrsim 0.14$
eV. This value sets an upper limit to our numerical calculation since we
will neglect Landau level mixing. In this work, we will thus restrict our
calculation to $\left\vert \Delta _{B}\right\vert \leq 0.1$ eV, a bias that
corresponds to an interlayer electric field of $E\approx 0.3$ V/nm.
Crossings between the non-interacting energy levels also occur within a
given Landau level. An example is shown in Fig. \ref{figure2} (b) where a
crossing between $E_{+,N+1,3}^{(0)}$ with spin up and $E_{-,N+1,3}^{(0)}$
with spin down at $B=10$ T occurs at $\Delta _{B}=0.014$ eV for $N=1.$Such
crossing is of course taken into account.

Defining $\nu _{\pm }$ as the filling factors for valleys $K_{\pm }$ and $%
\nu _{1,2}$ as the filling factors for layers $1$ and $2,$ we have for each
level (we now omit the indices $N,j$ to simplify the notation) 
\begin{eqnarray}
\nu _{1} &=&n_{1,+}\nu _{+}+n_{1,-}\nu _{-}, \\
\nu _{2} &=&n_{2,+}\nu _{+}+n_{2,-}\nu _{-},
\end{eqnarray}%
where the projectors 
\begin{eqnarray}
n_{1,+} &=&\left\vert c_{+,4}\right\vert ^{2}+\left\vert c_{+,3}\right\vert
^{2}, \\
n_{1,-} &=&\left\vert c_{-,1}\right\vert ^{2}+\left\vert c_{-,2}\right\vert
^{2}, \\
n_{2,+} &=&\left\vert c_{+,1}\right\vert ^{2}+\left\vert c_{+,2}\right\vert
^{2}, \\
n_{2,-} &=&\left\vert c_{-,3}\right\vert ^{2}+\left\vert c_{-,4}\right\vert
^{2}.
\end{eqnarray}%
Obviously, 
\begin{eqnarray}
n_{1} &=&n_{1,+}+n_{1,-}, \\
n_{2} &=&n_{2,+}+n_{2,-},
\end{eqnarray}%
and from the normalization condition $\sum_{i=1}^{i=4}\left\vert c_{\xi
,n,j,i}^{2}\right\vert =1$ for the eigenspinors%
\begin{equation}
n_{1,\pm }+n_{2,\pm }=1.
\end{equation}

The eigenspinor coefficients are related by 
\begin{equation}
c_{+,j}\left( \Delta _{B}\right) =\left[ c_{-,j}\left( -\Delta _{B}\right) %
\right] ^{\ast }
\end{equation}%
so that at zero bias 
\begin{equation}
n_{1,\mp }=n_{2,\pm }.
\end{equation}%
\qquad

\begin{figure}[tbph]
\includegraphics[scale=0.8]{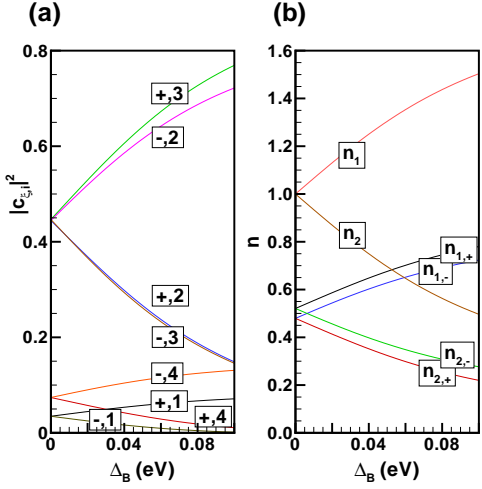}
\caption{(Color online) (a) Coefficients $\left\vert c_{\protect\xi %
,i}\right\vert ^{2}$ and (b) projectors $n_{1,\pm },n_{2,\pm },n_{1},n_{2}$
as a function of bias for $B=10$ T and\ Landau level $N=1$.}
\label{figure3}
\end{figure}

Figure \ref{figure3} shows the coefficients $\left\vert c_{\xi
,i}\right\vert ^{2}$ and the projectors $n_{1,\pm },n_{2,\pm },n_{1},n_{2}$
as a function of bias for $B=10$ T and $N=1.$ For a positive bias, layer $1$
has a higher energy than layer $2$ and Eq. (\ref{t3}) gives $%
c_{+,1},c_{+,2},c_{-,3},c_{-,4}\rightarrow 0$ at large positive bias so that
the electrons occupy layer 1 in this limit. This is consistent with the fact
that Landau levels $N>0,$ which correspond to antibonding states of the
layers, have higher energies than Landau levels $N<0$ which are bonding
states.

\section{HARTREE-FOCK\ HAMILTONIAN}

We now project the Hamiltonian into Landau level $N$ (with $N>0$ standing
for $N,j=3$ and $N<0$ standing for $N,j=2$). The Coulomb interaction in this
level is given by 
\begin{eqnarray}
V_{C} &=&\frac{1}{2}\sum\limits_{\xi ,\xi ^{\prime },\sigma ,\sigma ^{\prime
}}\int d\mathbf{u}\int d\mathbf{u}^{\prime }\Psi _{N,\xi ,\sigma }^{\dag
}\left( \mathbf{u}\right) \\
&&\times \Psi _{N,\xi ^{\prime },\sigma ^{\prime }}^{\dag }\left( \mathbf{u}%
^{\prime }\right) V\left( \mathbf{u}-\mathbf{u}^{\prime }\right) \Psi
_{N,\xi ^{\prime },\sigma ^{\prime }}\left( \mathbf{u}^{\prime }\right) \Psi
_{N,\xi ,\sigma }\left( \mathbf{u}\right) ,  \notag
\end{eqnarray}%
where terms that do not conserve the valley index have been neglected\cite%
{Goerbig}. The field operator $\Psi _{N,\xi ,\sigma }$ is defined by 
\begin{equation}
\Psi _{N,\xi ,\sigma }\left( \mathbf{u}\right) =\sum_{X}\psi _{\left\vert
N\right\vert +1,\xi ,X}^{\left( 0\right) }\left( \mathbf{u}\right) c_{N,\xi
,X,\sigma },
\end{equation}%
where $c_{N,\xi ,X,\sigma }$ annihilates an electron with quantum numbers $%
\xi ,N,\sigma $ where $\sigma =\pm 1$ is the spin index. The Coulomb
potential $V\left( \mathbf{u}-\mathbf{u}^{\prime }\right) $ is given by%
\begin{equation}
V\left( \mathbf{u}-\mathbf{u}^{\prime }\right) =\frac{1}{S}\sum_{\mathbf{q}}%
\frac{2\pi e^{2}}{\kappa q}e^{i\mathbf{q}\cdot \left( \mathbf{r}-\mathbf{r}%
^{\prime }\right) }e^{-q\left\vert z-z^{\prime }\right\vert },
\end{equation}%
where $\kappa $ is the dielectric constant of the substrate. When there is
no ambiguity, we will drop the index $N$ to simplify the notation.

We define the operator 
\begin{equation}
\rho _{\xi ,\xi ^{\prime }}^{\sigma ,\sigma ^{\prime }}\equiv \frac{1}{%
N_{\varphi }}\sum_{X}c_{\xi ,X,\sigma }^{\dagger }c_{\xi ^{\prime },X,\sigma
^{\prime }}
\end{equation}%
so that the set of average values $\left\{ \left\langle \rho _{\xi ,\xi
^{\prime }}^{\sigma ,\sigma ^{\prime }}\right\rangle \right\} $ can serve as
the order parameters of any uniform state of the C2DEG. The non uniform
states can be described by a space dependent analog of $\left\langle \rho
_{\xi ,\xi ^{\prime }}^{\sigma ,\sigma ^{\prime }}\right\rangle $\cite%
{Lambert}$.$ In this work, however, we consider only the uniform states.

The Hartree-Fock Hamiltonian $H_{HF}$ is obtained by making the Hartree and
Fock pairings of the field operators in $V_{C}$. The diverging part of the
Hartree interaction is cancelled by considering the interaction of the
electrons with the neutralizing positive background and so 
\begin{eqnarray}
H_{HF} &=&-N_{\varphi }\alpha _{d}\frac{\widetilde{\nu }^{2}}{4}
\label{hamil} \\
&&+N_{\varphi }\sum_{\xi ,\sigma }\left( \frac{\widetilde{\nu }}{2}\alpha
_{d}+E_{\xi ,\sigma }\right) \rho _{\xi ,\xi }^{\sigma ,\sigma }  \notag \\
&&-N_{\varphi }\sum_{\xi ,\zeta ,\sigma ,\sigma ^{\prime }}X^{\left( \xi
,\zeta \right) }\left\langle \rho _{\xi ,\zeta }^{\sigma ,\sigma ^{\prime
}}\right\rangle \rho _{\zeta ,\xi }^{\sigma ^{\prime },\sigma },  \notag
\end{eqnarray}%
where $\widetilde{\nu }$ is the filling factor of Landau level $N$ and we
have defined the constant%
\begin{equation}
\alpha _{d}=\frac{d}{\ell }\left( \frac{e^{2}}{\kappa \ell }\right)
\end{equation}%
as well as the renormalized single-particle energy $E_{\xi ,\sigma }$%
\begin{eqnarray}
E_{\xi ,\sigma } &=&E_{\xi }^{\left( 0\right) }-\frac{1}{2}\sigma \Delta _{Z}
\\
&&+\sum_{\zeta ,\sigma ^{\prime }}A^{\left( \xi ,\zeta \right) }\left\langle
\rho _{\zeta ,\zeta }^{\sigma ^{\prime },\sigma ^{\prime }}\right\rangle , 
\notag
\end{eqnarray}%
where $A_{N}^{\left( \xi ,\zeta \right) }$ is the non-diverging part of the
Hartree interaction $H^{\left( \xi ,\zeta \right) }\left( 0\right) $ (see
below) i.e. 
\begin{eqnarray}
A^{\left( +,+\right) } &=&-2\alpha _{d}n_{2,+}n_{1,+}, \\
A^{\left( +,-\right) } &=&A^{\left( -,+\right) }=-\alpha _{d}\left(
n_{2,+}n_{1,-}+n_{1,+}n_{2,-}\right) , \\
A^{\left( -,-\right) } &=&-2\alpha _{d}n_{1,-}n_{2,-}.
\end{eqnarray}%
In graphene, the Zeeman coupling $\Delta _{Z}=g\mu _{B}B=1.158\times 10^{-4}$
$B$ (T) eV.

The Hartree and Fock interactions are given by%
\begin{eqnarray}
H^{\left( \xi ,\xi ^{\prime }\right) }\left( x\right) &=&\left( \frac{e^{2}}{%
\kappa \ell }\right) \frac{V^{\left( \xi ,\xi ^{\prime }\right) }\left(
x\right) }{x}, \\
X^{\left( \xi ,\xi ^{\prime }\right) } &=&\left( \frac{e^{2}}{\kappa \ell }%
\right) \int_{0}^{\infty }dxV^{\left( \xi ,\xi ^{\prime }\right) }\left(
x\right) ,
\end{eqnarray}%
where 
\begin{align}
& V^{\left( \xi ,\xi \right) }\left( x\right) \\
& =G_{1}^{\left( \xi \right) }\left( x\right) G_{1}^{\left( \xi \right)
}\left( x\right) +G_{2}^{\left( \xi \right) }\left( x\right) G_{2}^{\left(
\xi \right) }\left( x\right)  \notag \\
& +2e^{-xd/\ell }G_{1}^{\left( \xi \right) }\left( x\right) G_{2}^{\left(
\xi \right) }\left( x\right) ,  \notag
\end{align}%
\begin{eqnarray}
&&V^{\left( \xi ,\overline{\xi }\right) }\left( x\right) \\
&=&G_{1}^{\left( \xi \right) }\left( x\right) G_{2}^{\left( \overline{\xi }%
\right) }\left( x\right) +G_{2}^{\left( \xi \right) }\left( x\right)
G_{1}^{\left( \overline{\xi }\right) }\left( x\right)  \notag \\
&&+e^{-xd/\ell }\left[ G_{1}^{\left( \xi \right) }\left( x\right)
G_{1}^{\left( \overline{\xi }\right) }\left( x\right) +G_{2}^{\left( \xi
\right) }\left( x\right) G_{2}^{\left( \overline{\xi }\right) }\left(
x\right) \right] ,  \notag
\end{eqnarray}%
where $\overline{\xi }=-\xi $ and the functions 
\begin{eqnarray}
&&G_{1}^{\left( \xi \right) }\left( x\right) \\
&=&e^{\frac{-x^{2}}{4}}\left[ \left\vert c_{\xi ,1}\right\vert
^{2}L_{N-1}\left( \frac{x^{2}}{2}\right) +\left\vert c_{\xi ,2}\right\vert
^{2}L_{N}\left( \frac{x^{2}}{2}\right) \right] ,  \notag \\
&&G_{2}^{\left( \xi \right) }\left( x\right) \\
&=&e^{\frac{-x^{2}}{4}}\left[ \left\vert c_{\xi ,3}\right\vert
^{2}L_{N-2}\left( \frac{x^{2}}{2}\right) +\left\vert c_{\xi ,4}\right\vert
^{2}L_{N-1}\left( \frac{x^{2}}{2}\right) \right] ,  \notag
\end{eqnarray}%
where $L_{N}\left( x\right) $ is a Laguerre polynomial with $L_{N<0}\left(
x\right) =0.$

\section{CALCULATION OF THE ORDER PARAMETERS}

The order parameters of a given phase can be computed by defining the
single-particle Green's function 
\begin{equation}
G_{\sigma ,\sigma ^{\prime }}^{\xi ,\xi ^{\prime }}\left( X,\tau \right)
=-\left\langle T_{\tau }c_{\xi ,\sigma ,X}\left( \tau \right) c_{\xi
^{\prime },\sigma ^{\prime },X}^{\dagger }\left( 0\right) \right\rangle ,
\end{equation}%
where $T_{\tau }$ is the imaginary time ordering operator.

If we define the Fourier transform of the single-particle Green's function
by 
\begin{equation}
G_{\sigma ,\sigma ^{\prime }}^{\xi ,\xi ^{\prime }}\left( \tau \right) =%
\frac{1}{N_{\phi }}\sum_{X}G_{\sigma ,\sigma ^{\prime }}^{\xi ,\xi ^{\prime
}}\left( X,\tau \right) ,
\end{equation}%
then the order parameters can be obtained with the relation%
\begin{equation}
\left\langle \rho _{\sigma ^{\prime },\sigma }^{\xi ^{\prime },\xi
}\right\rangle =G_{\sigma ,\sigma ^{\prime }}^{\xi ,\xi ^{\prime }}\left(
\tau =0^{-}\right) .
\end{equation}

The Green's function itself is obtained by solving the Hartree-Fock equation
of motion which is given by

\begin{eqnarray}
&&\left[ \hslash i\omega _{n}-\left( E_{\xi ,\sigma }-\mu \right) \right]
G_{\sigma ,\sigma ^{\prime }}^{\xi ,\xi ^{\prime }}\left( \omega _{n}\right)
=\hslash \delta _{\xi ,\xi ^{\prime }}\delta _{\sigma ,\sigma ^{\prime },}
\label{eqhfa} \\
&&-\sum_{\zeta ,\sigma ^{\prime \prime }}X^{\left( \zeta ,\xi \right)
}\left\langle \rho _{\sigma ^{\prime \prime },\sigma }^{\zeta ,\xi
}\right\rangle G_{\sigma ^{\prime \prime },\sigma ^{\prime }}^{\zeta ,\xi
^{\prime }}\left( \omega _{n}\right) ,  \notag
\end{eqnarray}%
where $\omega _{n}$ is a fermionic Matsubara frequency$.$ At $T=0$ K, it is
easy to show that the order parameters satisfy the sum rules%
\begin{equation}
\sum_{\sigma ^{\prime },\xi ^{\prime }}\left\vert \left\langle \rho _{\sigma
,\sigma ^{\prime }}^{\xi ,\xi ^{\prime }}\right\rangle \right\vert
^{2}=\left\langle \rho _{\sigma ,\sigma }^{\xi ,\xi }\right\rangle ,
\label{sumrule}
\end{equation}%
where $\left\langle \rho _{\sigma ,\sigma }^{\xi ,\xi }\right\rangle =\nu
_{\xi ,\sigma }$ is the filling factor of level $\left( \xi ,\sigma \right)
. $

Equation (\ref{eqhfa}) can easily be put in a matrix form%
\begin{equation}
\left[ I\left( \omega +i\delta \right) -F\right] G\left( \omega _{n}\right)
=B  \label{form}
\end{equation}%
and solved numerically. Because it is a self-consistent equation, it needs
to be solved iteratively starting with an initial set of order parameters.
The method has been described previously\cite{Coteprb91}. Once the order
parameters are found, the ground state energy per electron is given by%
\begin{eqnarray}
\frac{E}{N_{e}} &=&\frac{1}{4}\alpha _{d}\widetilde{\nu }+\frac{1}{%
\widetilde{\nu }}\sum_{\xi ,\sigma }\widetilde{E}_{\xi ,\sigma }\left\langle
\rho _{\xi ,\xi }^{\sigma ,\sigma }\right\rangle \\
&&-\frac{1}{2\widetilde{\nu }}\sum_{\xi ,\xi ^{\prime },\sigma ,\sigma
^{\prime }}X^{\left( \xi ,\xi ^{\prime }\right) }\left\vert \left\langle
\rho _{\xi ,\xi ^{\prime }}^{\sigma ,\sigma ^{\prime }}\right\rangle
\right\vert ^{2}.  \notag
\end{eqnarray}%
where%
\begin{equation}
\widetilde{E}_{\xi ,\sigma }=E_{\xi }^{\left( 0\right) }-\frac{1}{2}\sigma
\Delta _{Z}+\frac{1}{2}\sum_{\zeta ,\sigma ^{\prime }}A^{\left( \xi ,\zeta
\right) }\left\langle \rho _{\zeta ,\zeta }^{\sigma ^{\prime },\sigma
^{\prime }}\right\rangle .
\end{equation}

\section{PSEUDOSPIN DESCRIPTION}

It is useful at this point to introduce the valley pseudospin. We do this by
associating the up and down states of the valley pseudospin with the $K_{+}$
and $K_{-}$ valley states. We define the super index $i=1,2,3,4$ to denote
the four states $\left( \xi ,\sigma \right) :$ 
\begin{eqnarray}
\left( K_{+},+\right) &\rightarrow &1, \\
\left( K_{+},-\right) &\rightarrow &2, \\
\left( K_{-},+\right) &\rightarrow &3, \\
\left( K_{-},-\right) &\rightarrow &4.
\end{eqnarray}%
When $\widetilde{\nu }=1$ and all electrons are in the $K_{+}\left(
K_{-}\right) $ valleys, the pseudospin $P_{z}=+\frac{1}{2}\left( -\frac{1}{2}%
\right) .$ Valley coherence leads to a finite value of $P_{x}$ and $P_{y}.$
The total valley pseudospin is given by the sum of the valley pseudospin for
each spin component i.e. $\mathbf{P}=\mathbf{P}_{+}+\mathbf{P}_{-}$ with: 
\begin{eqnarray}
P_{x,+}+iP_{y,+} &=&\left\langle \rho _{1,3}\right\rangle , \\
P_{z,+} &=&\frac{1}{2}\left[ \left\langle \rho _{1,1}\right\rangle
-\left\langle \rho _{3,3}\right\rangle \right] , \\
P_{x,-}+iP_{y,-} &=&\left\langle \rho _{2,4}\right\rangle , \\
P_{z,-} &=&\frac{1}{2}\left[ \left\langle \rho _{2,2}\right\rangle
-\left\langle \rho _{4,4}\right\rangle \right] .
\end{eqnarray}%
\newline
For the real spin, the total spin $\mathbf{S}=\mathbf{S}_{+}+\mathbf{S}_{-}$
is given by the sum of the spin in each valley: 
\begin{eqnarray}
S_{x,+}+iS_{y,+} &=&\left\langle \rho _{1,2}\right\rangle , \\
S_{z,+} &=&\frac{1}{2}\left[ \left\langle \rho _{1,1}\right\rangle
-\left\langle \rho _{2,2}\right\rangle \right] , \\
S_{x,-}+iS_{y,-} &=&\left\langle \rho _{3,4}\right\rangle , \\
S_{z,-} &=&\frac{1}{2}\left[ \left\langle \rho _{3,3}\right\rangle
-\left\langle \rho _{4,4}\right\rangle \right] .
\end{eqnarray}%
Finally, the total filling factor of level $N$ is given by $\widetilde{\nu }%
=\nu _{+}+\nu _{-}$ where the filling factor for each valley is:

\begin{eqnarray}
\nu _{+} &=&\left\langle \rho _{1,1}\right\rangle +\left\langle \rho
_{2,2}\right\rangle , \\
\nu _{-} &=&\left\langle \rho _{3,3}\right\rangle +\left\langle \rho
_{4,4}\right\rangle .
\end{eqnarray}

The Hartree-Fock energy per electron can be written in terms of these fields
(which are not all independent variables) and the two order parameters $%
\left\langle \rho _{1,4}\right\rangle ,\left\langle \rho _{2,3}\right\rangle
.\mathbf{\ }$Note that the operators $\rho _{1,4},\rho _{2,3}$ flip both the
spin and the valley pseudospin.

\begin{eqnarray}
&&\frac{E}{N_{e}}  \label{hhfa} \\
&=&\frac{1}{2}\left( \Lambda _{+}^{\left( 0\right) }+\frac{\alpha _{d}}{2}%
\widetilde{\nu }-\frac{\alpha _{d}}{2}n_{1}n_{2}\widetilde{\nu }-\frac{%
\Lambda _{\rho ,\rho }}{8}\widetilde{\nu }\right)  \notag \\
&&-\frac{1}{\widetilde{\nu }}\Delta _{Z}S_{z}  \notag \\
&&+\frac{1}{\widetilde{\nu }}\left( \Lambda _{-}^{\left( 0\right) }-%
\widetilde{\nu }\alpha _{d}\left( n_{1,+}n_{2,+}-n_{1,-}n_{2,-}\right) -%
\frac{\widetilde{\nu }}{4}\Lambda _{\rho ,z}\right) P_{z}  \notag \\
&&-\frac{1}{\widetilde{\nu }}\left[ \alpha _{d}\left( n_{1,+}-n_{1,-}\right)
\left( n_{2,+}-n_{2,-}\right) +\frac{\Lambda _{z,z}}{4}\right] P_{z}^{2} 
\notag \\
&&-\frac{1}{\widetilde{\nu }}\left( X^{\left( +,+\right) }\left\vert \mathbf{%
S}_{+}\right\vert ^{2}+X^{\left( -,-\right) }\left\vert \mathbf{S}%
_{-}\right\vert ^{2}\right)  \notag \\
&&-\frac{1}{\widetilde{\nu }}X^{\left( +,-\right) }\left( \left\vert \mathbf{%
P}_{\bot ,+}\right\vert ^{2}+\left\vert \mathbf{P}_{\bot ,-}\right\vert
^{2}\right)  \notag \\
&&-\frac{1}{\widetilde{\nu }}X^{\left( +,-\right) }\left[ \left\vert
\left\langle \rho _{1,4}\right\rangle \right\vert ^{2}+\left\vert
\left\langle \rho _{2,3}\right\rangle \right\vert ^{2}\right] ,  \notag
\end{eqnarray}%
where we have defined the interactions%
\begin{eqnarray}
\Lambda _{\rho ,\rho } &=&\Lambda _{z,z}=X^{\left( +,+\right) }+X^{\left(
-,-\right) }, \\
\Lambda _{\rho ,z} &=&X^{\left( +,+\right) }-X^{\left( -,-\right) }
\end{eqnarray}%
and%
\begin{equation}
\Lambda _{\pm }^{\left( 0\right) }=E_{+}^{\left( 0\right) }\pm E_{-}^{\left(
0\right) }.
\end{equation}

We remark that the sum of the terms with $\alpha _{d}$ in Eq. (\ref{hhfa})
gives%
\begin{equation}
E_{C}=\frac{1}{4\widetilde{\nu }}\alpha _{d}\left( \rho _{1}-\rho
_{2}\right) ^{2}
\end{equation}%
which is just the capacitive energy of the graphene bilayer.

In pseudospin language, the four sum rules of Eq. (\ref{sumrule}) can be
added together to give%
\begin{eqnarray}
&&\frac{1}{4}\left\vert \widetilde{\nu }\right\vert ^{2}+\left\vert
P_{z}\right\vert ^{2}+2\left\vert P_{\bot ,+}\right\vert ^{2}+2\left\vert
P_{\bot ,-}\right\vert ^{2}  \label{sumrule3} \\
&&+2\left\vert \mathbf{S}_{+}\right\vert ^{2}+2\left\vert \mathbf{S}%
_{-}\right\vert ^{2}+2\left\vert \left\langle \rho _{1,4}\right\rangle
\right\vert ^{2}+2\left\vert \left\langle \rho _{2,3}\right\rangle
\right\vert ^{2}  \notag \\
&=&\widetilde{\nu }  \notag
\end{eqnarray}

\section{PHASE DIAGRAM FOR $\protect\widetilde{\protect\nu }=1$}

The Hartree-Fock formalism described above can easily be generalized to
study non-uniform states\cite{Fouquet}. The order parameters are then
wave-vector dependent. We have checked numerically that no square or
triangular Wigner crystals with or without spin/valley pseudospin texture is
possible at integer fillings. A crystal seed given to the numerical code for
solving the Hartree-Fock equations of motion always iterates to a uniform
state. A helical state where the layer pseudospin rotates along one
direction of space\cite{Fouquet} is possible but its energy is higher than
the uniform ground states discussed below.

At a quarter filling of a Landau level ($\widetilde{\nu }=1$), our numerical
calculation for a homogeneous state shows that the ground state is always
spin polarized i.e. $S_{z}=\frac{1}{2}$ and $\left\langle \rho
_{1,4}\right\rangle =\left\langle \rho _{2,3}\right\rangle =0.$ The
Hartree-Fock energy per electron thus simplifies to 
\begin{equation}
\frac{E}{N_{e}}=C-\left[ \Lambda _{z}P_{z,+}+J_{z}P_{z,+}^{2}+J_{\bot
}P_{\bot ,+}^{2}\right] ,  \label{pseudoferro}
\end{equation}%
where we have defined the constant 
\begin{eqnarray}
C &=&\frac{1}{4}\alpha _{d}+\frac{1}{2}\Lambda _{+}^{\left( 0\right) }-\frac{%
1}{2}\Delta _{Z} \\
&&-\frac{1}{4}\alpha _{d}n_{1}n_{2}-\frac{1}{8}\Lambda _{z,z},  \notag
\end{eqnarray}%
the bias term%
\begin{eqnarray}
\Lambda _{z} &=&\left( E_{-}^{0}-\frac{1}{2}X^{\left( -,-\right) }-\alpha
_{d}n_{1,-}n_{2,-}\right)  \label{lz} \\
&&-\left( E_{+}^{0}-\frac{1}{2}X^{\left( +,+\right) }-\alpha
_{d}n_{1,+}n_{2,+}\right) ,  \notag
\end{eqnarray}%
and the effective exchange interactions

\begin{eqnarray}
J_{z} &=&\alpha _{d}\left( n_{1,+}-n_{1,-}\right) \left(
n_{2,+}-n_{2,-}\right) +\frac{1}{2}\Lambda _{z,z},  \label{jzp} \\
J_{\bot } &=&X^{\left( +,-\right) }.  \label{jperpp}
\end{eqnarray}

\begin{figure}[tbph]
\includegraphics[scale=0.8]{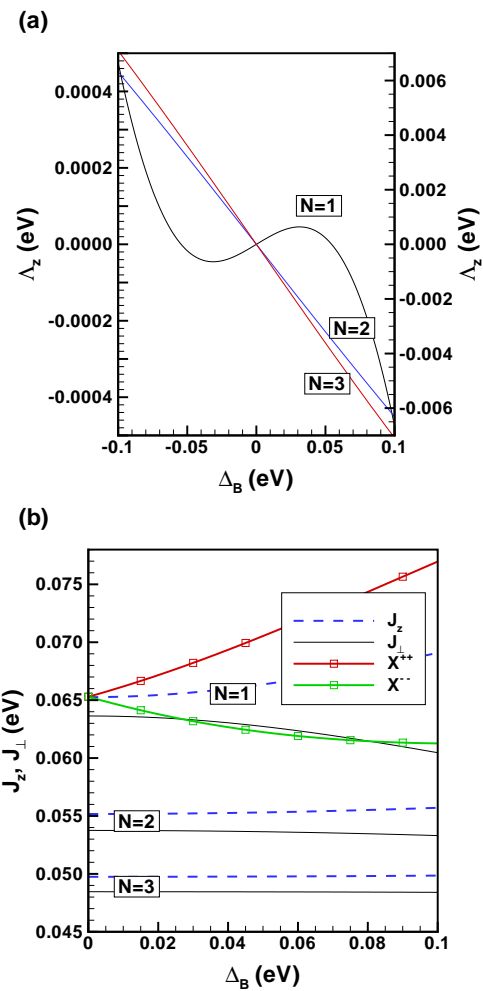}
\caption{(Color online) Effective interactions (a) $\Lambda _{z}$ and (b) $%
J_{z},J_{\bot }$ of the Ising quantum Hall valley-ferromagnet as a function
of the applied bias for $\protect\widetilde{\protect\nu }=1$ and $N=1,2,3$
at $B=16.5$ T, $\protect\kappa =2.5.$ The interactions $X^{++}$ and $X^{--}$
are also plotted for $N=1.$ In (a) the left axis is for $N=1$ and the right
axis for $N=2,3.$}
\label{figure4}
\end{figure}

The interactions $\Lambda _{z},J_{z}$ and $J_{\bot }$ are plotted in Fig. %
\ref{figure4} as functions of the bias for two values of $\kappa $ and for $%
B=16.5$ T. A change in bias or magnetic field modifies the coefficients of
the eigenspinors in Eqs. (\ref{psi1}),(\ref{psi2}). This, in turn modifies
the exchange interactions that enter in the definition of the interactions $%
\Lambda _{z},J_{z}$ and $J_{\bot }$. This modification of the interactions
with the applied bias did not occur in previous studies of the C2DEG in
Landau level $N=0.$ The reason is that the two-component model - which is a
good approximation for $N=0$ - was used\cite{Lambert} and, in this
simplified model, there is only one non zero component in the spinor for $%
n=0 $ and $n=1$ which is, of course, independent of bias.

Eq. (\ref{sumrule3}) gives $\left\vert \mathbf{P}_{+}\right\vert =\frac{1}{2}
$ so that we can write $\mathbf{P}_{+}$ in spherical coordinates to get 
\begin{eqnarray}
\frac{E}{N_{e}} &=&C-\frac{J_{\bot }}{4}-\frac{1}{2}\Lambda _{z}\cos \theta
\\
&&-\frac{1}{4}\left( J_{z}-J_{\bot }\right) \cos ^{2}\theta ,  \notag
\end{eqnarray}%
where $\theta $ is the angle between $\mathbf{P}_{+}$ and the $z$ axis. We
consider three cases of interest:

Case 1. In the artificial case where $\Delta _{B}=0$ and $d=0,$ the energies 
$E_{+}^{0}=E_{-}^{0}$ and $X^{\left( +,+\right) }=X^{\left( -,-\right)
}=X^{\left( +,-\right) }$ so that $\Lambda _{z}=0$ and $J_{z}=J_{\bot }.$
The C2DEG is a valley QHF with full SU(2) symmetry in the valley pseudospin
space.

Case 2. When $\Delta _{B}=0$ but $d\neq 0,$ the energies $%
E_{+}^{0}=E_{-}^{0} $ but $X^{\left( +,+\right) }=X^{\left( -,-\right) }\neq
X^{\left( +,-\right) }.$ The coefficients of the eigenspinors are related by 
$c_{+,N,j,i}=c_{-,N,j,i}^{\ast }.$ This implies that $n_{2,+}=n_{1,-}$ and $%
n_{1,+}=n_{2,-}$ so that $\Lambda _{z}=0.$ Since Fig. \ref{figure4} (b)
shows that $J_{z}>J_{\bot }$, the energy is minimized when $\sin \theta =0$
and there are two equivalent ground states corresponding to $P_{z,+}=\pm 
\frac{1}{2}.$ In each of these ground states, there is a charge imbalance
given by%
\begin{equation}
\rho _{1}-\rho _{2}=2P_{z,+}\left( n_{1,+}-n_{2,+}\right) .  \label{virtue}
\end{equation}%
The C2DEG can thus be described as an Ising QHF. (As Eq. (\ref{virtue})
shows, filled Landau levels do not contribute to the charge imbalance.)\ For 
$B=10$ T,$\kappa =2.5$ and $N=1,2,3,$ the charge imbalance is $\rho
_{1}-\rho _{2}\approx $ $0.05.$

Case 3. When $\Delta _{B}\neq 0$ and $d\neq 0,$ the C2DEG is an Ising QHF
but the sign of $P_{z,+}$ is now fixed by the bias term $\Lambda _{z}$
(which, in view of Eq. (\ref{pseudoferro}) is simply the difference in the
energy per electron between the two phases $P_{z,+}=\pm \frac{1}{2}.$) For $%
\Lambda _{z}>0$, $P_{z,+}=+\frac{1}{2}$ while for $\Lambda _{z}<0$, $%
P_{z,+}=-\frac{1}{2}.$ Alternatively, the energy of these two ground states
can be written as%
\begin{equation}
\frac{E_{+}}{N_{e}}=E_{+}^{0}-\frac{1}{2}\Delta _{z}-\alpha
_{d}n_{1,+}n_{2,+}-\frac{1}{2}X^{\left( +,+\right) },
\end{equation}%
\begin{equation}
\frac{E_{-}}{N_{e}}=E_{-}^{0}-\frac{1}{2}\Delta _{z}-\alpha
_{d}n_{1,-}n_{2,-}-\frac{1}{2}X^{\left( -,-\right) }.
\end{equation}

Figure \ref{figure4} (a) shows that, for Landau level $N=1,\kappa =2.5$ and $%
B=16.5$ T, the ground state has $P_{z,+}=+\frac{1}{2}$ for $\Delta
_{B}<\Delta _{B}^{\left( c\right) }$ and $P_{z,+}=-\frac{1}{2}$ for $\Delta
_{B}>\Delta _{B}^{\left( c\right) }$ eV where the critical bias is $\Delta
_{B}^{\left( c\right) }=0.054$ eV. This pseudospin-flip transition does not
originate from Landau level crossing as is often the case with Ising QHF
since there is no crossing between the non-interacting levels $\left(
K_{+},+\right) $ and $\left( K_{-},+\right) $ in the energy spectrum (see
Fig. \ref{figure2} (b)). Instead, the transition is exchange-energy driven.
Although $E_{+,+}>E_{-,+},$ the phase with $P_{z,+}=+\frac{1}{2}$ has all
electrons in level $\left( K_{+},+\right) $ because the increase in kinetic
energy in this phase is more than compensated by the diminution of the
exchange energy since $X^{\left( +,+\right) }>X^{\left( -,-\right) }$ (see
Fig. \ref{figure4} (b)). The exchange energy is very sensitive to the
relative distribution of the amplitude of the electronic wave function on
the different Landau level orbitals which is given by the four-component
spinors in Eqs. (\ref{psi1}),(\ref{psi2}).

\begin{figure}[tbph]
\includegraphics[scale=0.8]{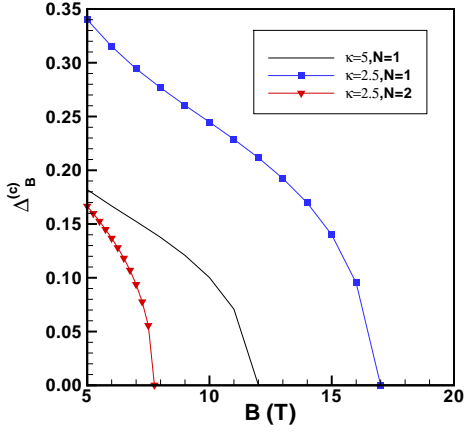}
\caption{(Color online) Critical bias $\Delta _{B}^{\left( c\right) }$ as a
function of the magnetic field in Landau levels $N=1,2$ at $\protect%
\widetilde{\protect\nu }=1$ for $\protect\kappa =2.5$ and $\protect\kappa %
=5. $ The region below (above)\ each curve is valley polarized with $P_{z}=+%
\frac{1}{2}\left( -\frac{1}{2}\right) .$ Biases above $\Delta _{B}^{\left(
c\right) }=0.10$ eV are outside the limit of validity of the model.}
\label{figure5}
\end{figure}

The critical bias $\Delta _{B}^{\left( c\right) }$ depends sensitively on
the value of the dielectric constant $\kappa .$ Its value is decreased by
increasing $\kappa $ as shown in Fig. \ref{figure5}. This figure also shows
that the phase transition line is shifted to lower magnetic fields when $N$
is increased. For the range of bias shown in Fig. \ref{figure4}, the ground
state has $P_{z,+}=-\frac{1}{2}$ for $N>2$ and there is no phase transition
in these levels. For levels $N=-2,-3,-4,$ the ground state has $P_{z,+}=-%
\frac{1}{2}$ at all bias, with the exception of $\Delta _{B}=0,$ in the
phase space shown in Fig. \ref{figure5}.

\begin{figure}[tbph]
\includegraphics[scale=0.8]{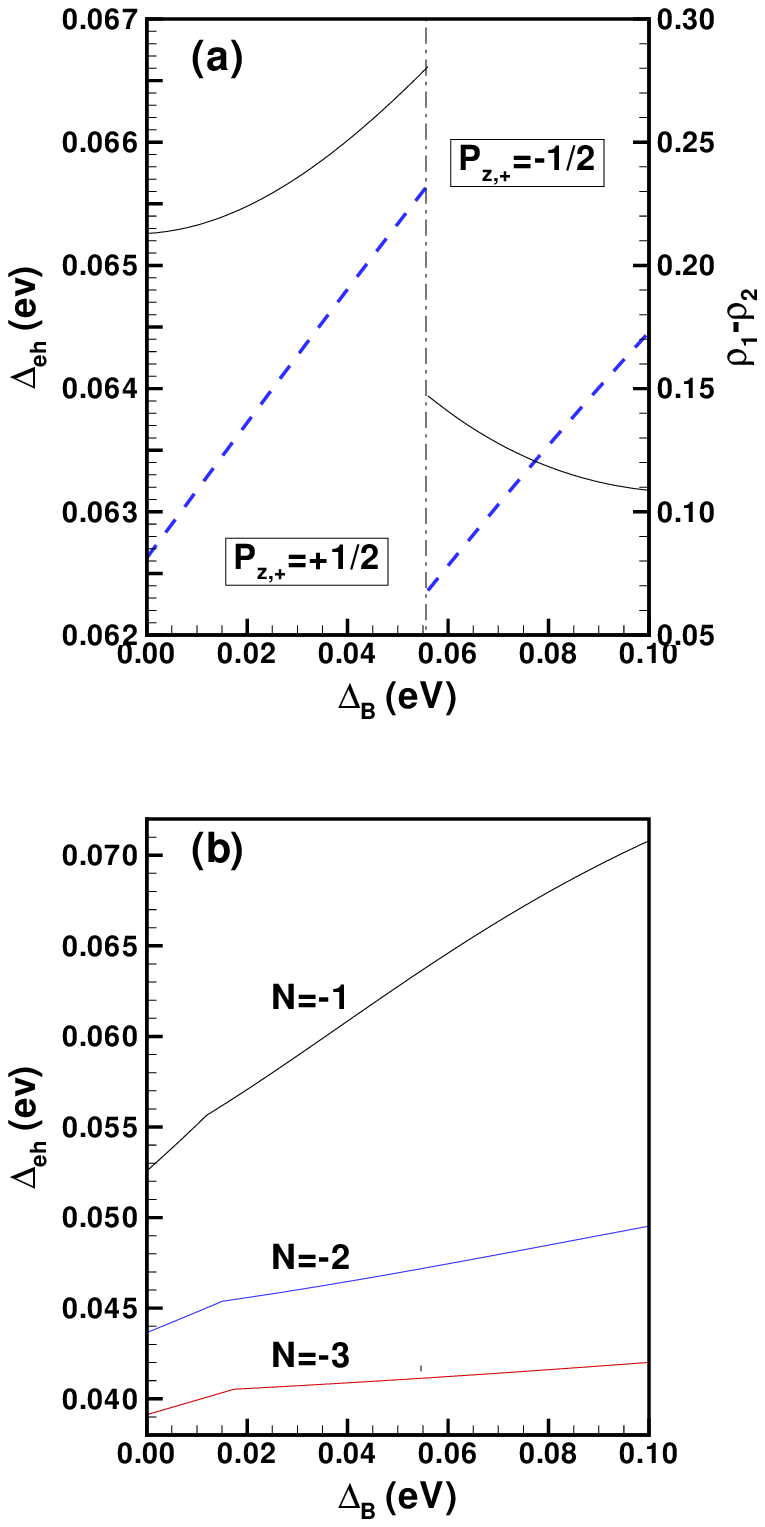}
\caption{(Color online) (a) Hartree-Fock gap $\Delta _{eh}$ (left axis,
black curve) and charge imbalance $\protect\rho _{1}-\protect\rho _{2}$
(right axis, blue dashed curve) plotted as a function of the applied bias
for $N=1,B=16.5$ T, $\protect\widetilde{\protect\nu }=1,\protect\kappa =2.5$%
. Both functions are discontinuous at the transition indicated by the
dashed-dotted line. (b) Hartree-Fock gap vs bias for $N=-1,-2,-3$ at $B=10$
T and for $\protect\kappa =2.5.$}
\label{figure6}
\end{figure}

Figure \ref{figure6} shows the behavior of the transport gap $\Delta _{eh}$
and density imbalance as the bias is varied for $B=16.5$ T, $\kappa =2.5$
and $N=1$ (the parameters of Fig. \ref{figure4}) The gap and the charge
imbalance have a jump at the transition between the two phases. The
Hartree-Fock or transport gap is defined as the energy to create an
infinitely separated (i.e. non interacting) electron-hole pair. It
corresponds to the difference in energy between the lowest unoccupied and
highest occupied single-particle Hartree-Fock levels. The energy of these
levels are given by the eigenvalues of the matrix $F$ in Eq. (\ref{form}).
Figure \ref{figure6} also shows the behavior of the gap for $N=-1,-2,-3,B=10$
T and $\kappa =2.5.$ There is no phase transition at finite bias in levels $%
N<0$ so that the discontinuity in the slope of the gap occurs because of a
crossing between the second and third Hartree-Fock level given by the matrix 
$F.$ (There is however a transition exactly at zero bias since $\Lambda _{z}$
changes sign there.) The discontinuity in the slope is more pronounced at
higher magnetic field.

\section{PHASE DIAGRAM FOR $\protect\widetilde{\protect\nu }=3$}

At $\widetilde{\nu }=3,$ the Hartree-Fock energy per electron can be written
as%
\begin{equation}
\frac{E}{N_{e}}=C^{\prime }-\frac{1}{3}\left[ \Lambda _{z}^{\prime
}P_{z,-}+J_{z}P_{z,-}^{2}+J_{\bot }P_{\bot ,-}^{2}\right] ,
\end{equation}%
with $J_{z}$ and $J_{\bot }$ still defined by Eqs. (\ref{jzp}) and (\ref%
{jperpp}) but with the constant 
\begin{eqnarray}
C^{\prime } &=&\frac{1}{2}\left( E_{-}^{0}+E_{+}^{0}\right) -\frac{3}{4}%
\alpha _{d}n_{1}n_{2}+\frac{3}{4}\alpha _{d} \\
&&-\frac{5}{24}\Lambda _{z,z}-\frac{1}{6}\Delta _{Z},  \notag
\end{eqnarray}%
and the bias term 
\begin{eqnarray}
\Lambda _{z}^{\prime } &=&\left( E_{-}^{0}-\frac{1}{2}X^{\left( -,-\right)
}-3\alpha _{d}n_{1,-}n_{2,-}\right) \\
&&-\left( E_{+}^{0}-\frac{1}{2}X^{\left( +,+\right) }-3\alpha
_{d}n_{1,+}n_{2,+}\right) .  \notag
\end{eqnarray}

The electron-hole symmetry in Landau level $N$ is not perfect since $\Lambda
_{z}^{\prime }\neq \Lambda _{z}.$ However, because the terms $\alpha
_{d}n_{1,\pm }n_{2,\pm }$ are very small compared to $E_{+}^{0}$ and $%
X_{N}^{\left( \pm ,\pm \right) }$, $\Lambda _{z}^{\prime }\approx $ $\Lambda
_{z}$ and the phase diagram for $\widetilde{\nu }=3$ is quasi identical to
that for $\widetilde{\nu }=1$.

The C2DEG is thus again an Ising QHF for $\widetilde{\nu }=3$ but with $%
P_{z,+}$ replaced by $P_{z,-}$. At zero bias, $\Lambda _{z}^{\prime }=0$ and
the two ground states $P_{z,-}=\pm \frac{1}{2}$ are degenerate. The charge
imbalance is again given by Eq. (\ref{virtue}) but with $P_{z,+}$ replaced
by $P_{z,-}.$

\section{PHASE DIAGRAM AT $\protect\widetilde{\protect\nu }=2$}

For a uniform ground state at $\widetilde{\nu }=2$, numerical calculations
show that states with valley and/or spin coherence do not occur. The
Hartree-Fock energy per electron can thus be simplified to%
\begin{eqnarray}
\frac{E}{N_{e}} &=&D-\frac{1}{2}\Delta _{Z}S_{z}-J_{z}P_{z}^{^{2}} \\
&&-\Lambda _{z}P_{z}-J_{+}S_{z,+}^{2}-J_{-}S_{z,-}^{2},  \notag
\end{eqnarray}%
where the constant $D$ is given by 
\begin{equation}
D=\frac{1}{2}\left( E_{-}^{0}+E_{+}^{0}+\alpha _{d}-\alpha
_{d}n_{1}n_{2}\right) -\frac{1}{4}\left( J_{+}+J_{-}\right) ,
\end{equation}%
the effective Heisenberg exchange interactions are%
\begin{eqnarray}
J_{z} &=&\frac{1}{2}\alpha _{d}\left( n_{1,+}-n_{1,-}\right) \left(
n_{2,+}-n_{2,-}\right) \\
&&+\frac{1}{4}\left( J_{+}+J_{-}\right) ,  \notag \\
J_{\pm } &=&\frac{1}{2}X^{\left( \pm ,\pm \right) },
\end{eqnarray}%
and the bias term is%
\begin{eqnarray}
\Lambda _{z} &=&-\frac{1}{2}\left( E_{+}^{0}-E_{-}^{0}\right) +\alpha
_{d}\left( n_{1,+}n_{2,+}-n_{1,-}n_{2,-}\right) \\
&&+\frac{1}{2}\left( J_{+}-J_{-}\right) .  \notag
\end{eqnarray}

In the absence of coherence, the only possible states have $P_{z}=0,\pm 1$
and $S_{z,+},S_{z,-}=\pm \frac{1}{2}.$ However, the sum rule of Eq. (\ref%
{sumrule}), which can be rewritten as, 
\begin{equation}
\frac{1}{2}-\frac{1}{2}\left\vert P_{z}\right\vert
^{2}=S_{z,+}^{2}+S_{z,-}^{2}
\end{equation}%
permits only six combinations. Three of them, with $S_{z}=0,-1,$ must be
ruled out since they have higher energies than the state with $S_{z}=1.$ We
only need to compare the energies of the three following states to establish
the phase diagram:

\begin{itemize}
\item Phase 1 is spin polarized and valley unpolarized. It has $S_{z,\pm }=%
\frac{1}{2},P_{z}=0$ and energy%
\begin{equation}
\frac{E_{1}}{N_{e}}=D-\frac{1}{2}\Delta _{Z}-\frac{1}{4}\left(
J_{+}+J_{-}\right) .
\end{equation}

\item Phase 2 is spin unpolarized and valley polarized. It has $S_{z,\pm
}=0,P_{z}=+1$ and energy%
\begin{equation}
\frac{E_{2}}{N_{e}}=D-J_{z}-\Lambda _{z}.
\end{equation}

\item Phase 3 is spin unpolarized and valley polarized. It has $S_{z,\pm
}=0,P_{z}=-1$ and energy%
\begin{equation}
\frac{E_{3}}{N_{e}}=D-J_{z}+\Lambda _{z}.
\end{equation}
\end{itemize}

\begin{figure}[tbph]
\includegraphics[scale=0.8]{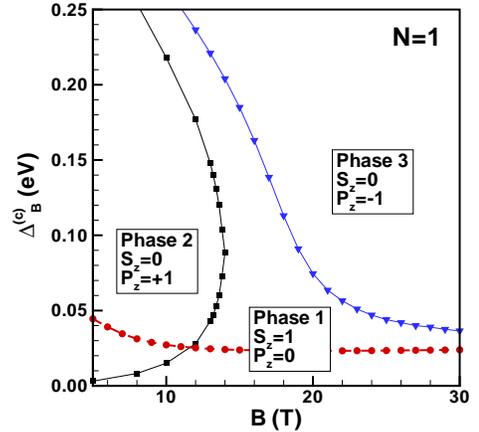}
\caption{(Color online) Phase diagram in Landau level $N=1$ for filling
factor $\protect\widetilde{\protect\nu }=2$ and dielectric constant $\protect%
\kappa =2.5.$ The red dashed line with the filled red circles is the
non-interacting result for the critical bias.}
\label{figure7}
\end{figure}

\begin{figure}[tbph]
\includegraphics[scale=0.8]{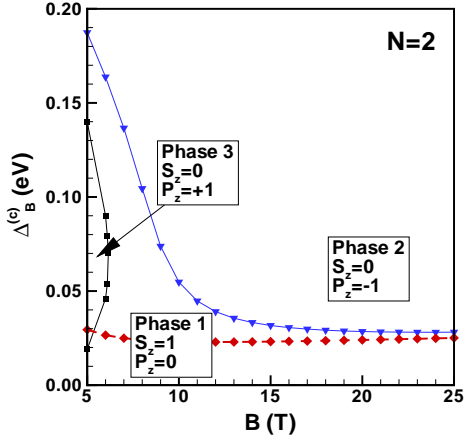}
\caption{(Color online) Phase diagram in Landau level $N=2$ for filling
factor $\protect\widetilde{\protect\nu }=2$ and dielectric constant $\protect%
\kappa =2.5.$ The red dashed line with the filled red circles is the
non-interacting result for the critical bias.}
\label{figure8}
\end{figure}

At zero bias, $\Lambda _{z}=0,J_{+}=J_{-}$ and $E_{2}=E_{3}.$ The energy $%
E_{1}<E_{2},E_{3}$ if the condition $\alpha _{d}\left(
n_{1,+}-n_{1,-}\right) ^{2}<\Delta _{Z}$ is satisfied, which is always the
case. Thus, the ground state is always spin polarized at zero bias. Figures %
\ref{figure7} and \ref{figure8} show the phase diagram for Landau levels $%
N=1 $ and $N=2$ with the dielectric constant $\kappa =2.5.$ The range of
bias in these figures is extended beyond the limit of validity of our model
in order to show the reentrant spin polarized phase transition that the
model would predict. When the bias is increased from zero, phase 1 can make
a transition to phase 2 or phase 3. In these figures, the black line with
the filled squares separates phase 2 on the left from phase 1 on the right
while the blue line with the filled triangles separate phase 1 on the left
from phase 3 on the right. Notice that the region corresponding to phase 2
is much smaller for $N=2$ than $N=1.$ For $N=2,$ in most of the phase space,
the transition is directly from phase 1 to phase 3.

The transition from phase 1 to phase 2 is exchange-energy driven, just as
the pseudospin-flip transition we discussed above for $\widetilde{\nu }=1,3$
was. It does not come from a level crossing. The transition from phase 1 to
phase 3, however, is what would be expected from the energy-level diagram of
Fig. \ref{figure2}. That is, level $\left( K_{+},+\right) $ crosses level $%
\left( K_{-},-\right) $ so that the occupied levels in the ground state are $%
\left( K_{+},+\right) ,\left( K_{-},+\right) $ in phase 1 and $\left(
K_{-},+\right) ,\left( K_{-},-\right) $ in phase 3. The energy of an
occupied level is however strongly modified by the exchange interaction and
so the phase transition does not occur at the value given by the crossing of
the non-interacting levels which is given by the red dashed line in Figs. %
\ref{figure7} and \ref{figure8} (this line separates phase 1 below from
phase 3 above). Comparing the non-interacting and the Hartree-Fock results,
one sees that the inclusion of the Coulomb exchange interaction radically
changes the phase diagram. This is less so for levels $N<0$ as we show below.

The hopping term $\gamma _{4}$ as well as $\delta _{0}$ that were included
in the tight-binding model lead to an electron-hole asymmetry. For this
reason, negative Landau levels must be considered separately. Figure \ref%
{figure9} shows the phase diagram for $N=-1,-2,-3$ and $\kappa =2.5.$ Phase $%
2$ is absent from the phase diagram and the predictions of the Hartree-Fock
theory is qualitatively the same as those of the non-interacting model
obtained from the crossing of the $\left( K_{+},+\right) $ and $\left(
K_{-},-\right) $ levels in the energy spectrum. (Note that for $N<0,$ the
levels disperse downward in energy instead of upward as in Fig. \ref{figure2}%
.) The phase diagram for the negative Landau level is obviously not as rich
as the one for the positive levels. For $N<0,$ the critical bias evaluated
in the absence of Coulomb interaction is bigger than the critical bias found
with the Hartree-Fock approximation. For $N>0,$ it is the other way around.

The transition from phase 1 to phase 3 is obtained in the HFA by solving the
equation $E_{1}=E_{3}$ i.e.

\begin{eqnarray}
E_{+}^{0}-E_{-}^{0} &=&\Delta _{Z}-\frac{1}{2}A^{\left( +,+\right) }+\frac{3%
}{2}A^{\left( -,-\right) } \\
&&-A^{\left( -,+\right) }+J_{+}-J_{-},  \notag
\end{eqnarray}%
while the non-interacting result is obtained from $E_{+}^{0}-E_{-}^{0}=%
\Delta _{Z}.$ The main correction to the non-interacting result comes from
the exchange term $J_{+}-J_{-}$.

\begin{figure}[tbph]
\includegraphics[scale=0.8]{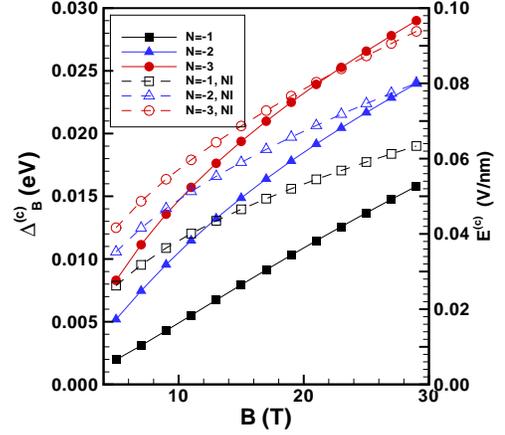}
\caption{(Color online) Critical bias $\Delta _{B}^{(c)}$ (left axis) and
critical electric field $E^{(c)}=\Delta _{B}/d$ (right axis) for the spin
polarized to spin unpolarized phase with electrons in valley $K_{-}$ for
filling factor $\protect\widetilde{\protect\nu }=2$ and dielectric constant $%
\protect\kappa =2.5$ in Landau levels $N=-1,-2,-3.$ The dashed lines give
the non-interacting results for $\Delta _{B}^{(c)}$ and $E^{(c)}.$}
\label{figure9}
\end{figure}

The transport gap $\Delta _{eh}$ and charge imbalance corresponding to Fig. %
\ref{figure7} for a magnetic field of $B=10$ T is plotted in Fig. \ref%
{figure10}. As for the pseudospin-flip transition discussed above for $%
\widetilde{\nu }=1,3,$ both quantities are discontinuous at the transition.
By contrast, Fig. \ref{figure11} shows that the transport gap is
discontinuous at the transition from phase 1 to phase 3 in Landau level $N=1$
for $B<24$ T but is continuous above $B^{\left( c\right) }=24$ T. The gap
closes progressively with $B.$ The same situation occurs for $N=2$ where the
gap is continuous above $B^{\left( c\right) }=11$ T. The closing of the gap
for $N=1,2$ occurs because of a crossing between the two lowest
single-particle Hartree-Fock levels. From the matrix $F$ in Eq. (\ref{form}%
), the energy of each level is given, in order of increasing energy for $%
B<B^{\left( c\right) },$ by%
\begin{eqnarray}
e_{1} &=&E_{-}^{0}-\frac{\Delta _{z}}{2}+A^{\left( -,+\right) }+A^{\left(
-,-\right) }-X^{\left( -,-\right) }, \\
e_{2} &=&E_{+}^{0}-\frac{\Delta _{z}}{2}+A^{\left( +,+\right) }+A^{\left(
+,-\right) }-X^{\left( +,+\right) }, \\
e_{3} &=&E_{-}^{0}+\frac{\Delta _{z}}{2}+A^{\left( -,+\right) }+A^{\left(
-,-\right) }, \\
e_{4} &=&E_{+}^{0}+\frac{\Delta _{z}}{2}+A^{\left( +,+\right) }+A^{\left(
+,-\right) },
\end{eqnarray}%
while for $B>B^{\left( c\right) },$ $e_{2}<e_{1}$. The electron-hole gap is
thus $\Delta _{eh}=e_{3}-e_{2}$ for $B<B^{\left( c\right) }$ and $\Delta
_{eh}=e_{3}-e_{1}$ for $B>B^{\left( c\right) }.$ Using the fact that $%
E_{1}=E_{3}$ at the transition, it is easy to show analytically that $\Delta
_{eh}=0$ for $B>B^{\left( c\right) }.$

The behavior of $\Delta _{eh}$ with bias for $N=-1,-2,-3$ is shown in Fig. %
\ref{figure12}. The gap has a downward cusp at the transition.

\begin{figure}[tbph]
\includegraphics[scale=0.8]{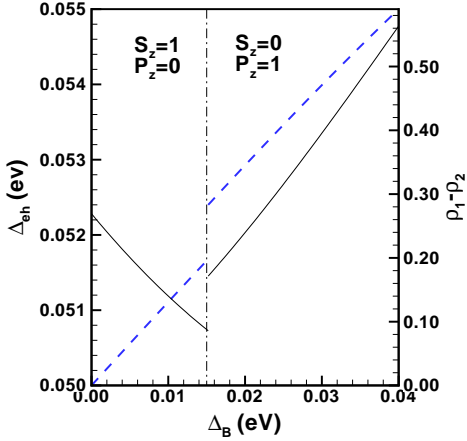}
\caption{(Color online) Hartree-Fock gap $\Delta _{eh}$ (left axis, full
line) and charge imbalance $\protect\rho _{1}-\protect\rho _{2}$ (right
axis, dashed line) vs applied bias for $\protect\widetilde{\protect\nu }%
=2,B=10$ T, $\protect\kappa =2.5$ and $N=1.$ The dashed-dotted line
indicates the critical bias for the phase transition.}
\label{figure10}
\end{figure}

\begin{figure}[tbph]
\includegraphics[scale=0.8]{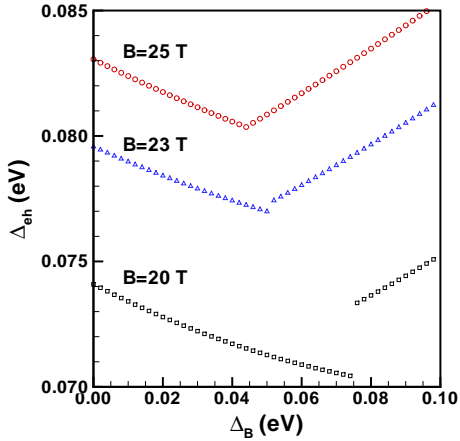}
\caption{(Color online) Hartree-Fock gap $\Delta _{eh}$ as a function of the
applied bias for $\protect\widetilde{\protect\nu }=2$ and $B=20,23,25$ T, $%
\protect\kappa =2.5,$ $N=1.$}
\label{figure11}
\end{figure}

\begin{figure}[tbph]
\includegraphics[scale=0.8]{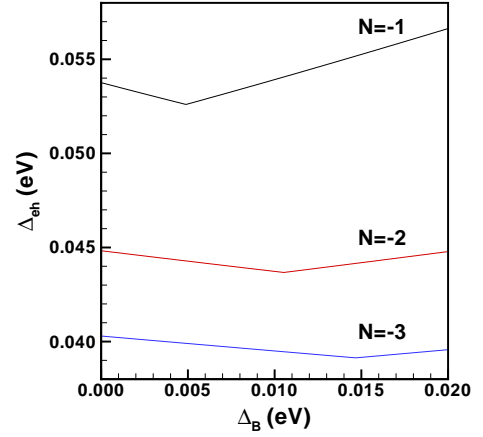}
\caption{(Color online) Hartree-Fock gap $\Delta _{eh}$ as a function of the
applied bias for $\protect\widetilde{\protect\nu }=2,$ $B=10$ T, $\protect%
\kappa =2.5,$ $N=-1,-2,-3.$ }
\label{figure12}
\end{figure}

\section{DISCUSSION\ AND\ CONCLUSION}

The experimental study of the C2DEG in BLG has so far been concentrated on
the phase diagram in Landau level $N=0$\cite{BLGExperiments} or to the
measurement of the transport gaps between higher Landau levels\cite%
{Exceptions}. In a recent publication\cite{Tutuc}, however, the QHFs that
emerge from the quartet of states in Landau levels $N<0$ are studied
experimentally (along with the states in $N=0$) using a double bilayer
graphene heterostructure\cite{Tutuc2}. The authors find quantum Hall states
at all integer filling factors, which undergo transitions as a function of
magnetic and transverse electric fields. At odd filling factors, the QHE\ is
absent at and near zero bias and reemerge at finite bias while at $%
\widetilde{\nu }=2$ there is a finite critical bias $\Delta _{B}^{\left(
c\right) }$around which the QHE is lost.

In a non-interacting picture for the energy levels, the transition at $%
\widetilde{\nu }=2$ is due to a crossing of the two sublevels $\left(
K_{+},+\right) $ and $\left( K_{-},-\right) $ and the ground state changes
from a spin polarized to a valley polarized state while the absence of the
QHE near zero bias at $\widetilde{\nu }=1$ and $\widetilde{\nu }=3$ is due
to the degeneracy of the states $\left( K_{+},+\right) $ and $\left(
K_{-},+\right) $ in the former case and $\left( K_{+},-\right) $ and $\left(
K_{-},-\right) $ in the latter (see Fig. \ref{figure2}). In this picture the
transport gap $\Delta _{eh}$ goes to zero at the level crossing (or
degeneracy point), the quantum Hall state is lost and the longitudinal
resistance $\rho _{xx}$ increases.

When Coulomb interaction is considered in the Hartree-Fock approximation for
Landau levels $N<0$, the gap is finite at all bias but has a downward cusp
at the transition between the spin polarized and the valley polarized state
at $\widetilde{\nu }=2$ (see Fig. \ref{figure12}) and near zero bias at
filling factors $\widetilde{\nu }=1,3$ (see Fig. \ref{figure6} (b)). If we
assume that, in the cusp region the Landau level broadening due to the
disorder is larger than the Hartree-Fock gap, then the QHE is lost in this
region and the phase transitions found in the HFA are consistent with the
experimental results. For this argument to hold, however, the broadening
must depend on the Landau level index. Note that the gaps calculated in the
Hartree-Fock approximation are exchange-enhanced and so larger than the
non-interacting gaps. In fact, they are of the same order than the gap
between Landau levels $N$ so that the applicability of no Landau-level
mixing approximation seems questionable. However, it was shown previously%
\cite{Gorbar} that static screening can reduce the size of the gaps
substantially so that the no mixing approximation can be justified. The
inclusion of screening corrections may well modify the phase diagrams
discussed in this paper however. As we have shown above, some of the phases
like that with $S_{z}=\frac{1}{2},P_{z}=\frac{1}{2}$ for $\widetilde{\nu }%
=1,3$ and phase 2 ($S_{z}=0,P_{z}=1$) for $\widetilde{\nu }=2$ are sensitive
to the value of $\kappa $ and thus to static screening. If these phase
disappears with screening, then the phase diagram for $N>0$ will look more
like that for $N<0.$ At the moment, there is no data for $N>0$ to which we
can compare our results.

In Fig. 3(d) of Ref. \onlinecite{Tutuc}, the critical bias $\Delta
_{B}^{\left( c\right) }\left( B\right) ,$ corresponding to observed
transition for $\widetilde{\nu }=2,$ is given for $N=0,-1,-2,-3.$ In our
terminology, this $\Delta _{B}^{\left( c\right) }\left( B\right) $
corresponds to the transition from phase 1 to phase 3 for which the phase
diagram is given in Fig. \ref{figure9}. Qualitatively, our results for $%
\Delta _{B}^{\left( c\right) }\left( B\right) $ agree well with experiment.
The critical bias (or critical electric field) increases almost linearly
with magnetic field at small field and it increases with Landau level index $%
\left\vert N\right\vert $ but more slowly as $\left\vert N\right\vert $
increases. Quantitatively, the comparison is more difficult because our
calculation does not include the disorder which is always present in a real
sample. The theoretical critical bias is about eight times smaller than the
experimental one depending on the level $N$. As for the slope of $\Delta
_{B}^{\left( c\right) }\left( B\right) $ with magnetic field, for $N=-1$ it
is $\approx 1.8$ mV$\cdot $nm$^{-1}\cdot $T$^{-1}$ while the experimental
result is larger and $\approx 9$ mV$\cdot $nm$^{-1}\cdot $T$^{-1}$. These
differences in the slope and value of the critical bias are similar in size
to those found between the theoretical\cite{Gorbar,Lambert} and experimental%
\cite{BLGExperiments} results for the spin polarized to layer-polarized
phase transition that occurs at filling factor $\nu =0$ in level $N=0$ in
BLG. Further study is necessary to understand the reason for this
discrepancy.

We have, in this work, concentrated our analysis on the uniform states in
the phase diagram. The formalism we developped can however be applied to the
study of non-uniform states such as charge-density-wave of crystals. We will
discuss these states elsewhere\cite{Domainwalls} together with the charged
excitations of the QHF states. Because of the Ising character of the QHF
states, the charged excitations can take the form of charged domain wall
loops (i.e. Skyrmions)\cite{Walls}. The transport gap $\Delta _{eh}$
computed in this paper can be modified if these topological excitations have
lower energy than the electron-hole pair excitations we considered in this
work.

\begin{acknowledgments}
R. C\^{o}t\'{e} was supported by a grant from the Natural Sciences and
Engineering Research Council of Canada (NSERC). Computer time was provided
by Calcul Qu\'{e}bec and Compute Canada. We thank Emanuel Tutuc and Kayoung
Lee for helpful discussions.
\end{acknowledgments}

\end{document}